\documentclass[prl,preprint,showpacs]{revtex4-1}
\usepackage{amsfonts}
\usepackage{amsmath}
\usepackage{amssymb}
\usepackage{lineno}
\usepackage{graphicx}%
\begin{document}

\linenumbers

\title{Spin-state crossover and hyperfine interactions of ferric iron
in MgSiO$_3$ perovskite}

\author{Han Hsu,$^{1}$ Peter Blaha,$^{2}$ Matteo Cococcioni,$^{1}$ and
Renata M. Wentzcovitch$^{1}$}

\affiliation{$^{1}$Department of Chemical Engineering and Materials
Science, University of Minnesota, Minneapolis, Minnesota, USA\\
$^2$Institute of Materials Chemistry, Vienna University of
Technology, A-1060 Vienna, Getreidemarkt 9/165-TC, Austria}

\date{\today}

\bigskip

\begin{abstract}
Using density functional theory plus Hubbard $U$ calculations, we
show that the ground state of (Mg,Fe)(Si,Fe)O$_3$ perovskite, the
major mineral phase in the Earth's lower mantle, has high-spin
ferric iron ($S=5/2$) at both dodecahedral (A) and octahedral (B)
sites. With increasing pressure, the B-site iron undergoes a
spin-state crossover to the low-spin state ($S=1/2$) between 40 and
70 GPa, while the A-site iron remains in the high-spin state. This
B-site spin-state crossover is accompanied by a noticeable volume
reduction and an increase in quadrupole splitting, consistent with
recent X-ray diffraction and M\"ossbauer spectroscopy measurements.
The anomalous volume reduction leads to a significant softening in
the bulk modulus during the crossover, suggesting a possible source
of seismic-velocity anomalies in the lower mantle.

\end{abstract}

\pacs{91.60.Pn, 76.80.+y, 91.60.Gf, 91.60.Fe}

\maketitle

The total electron spin ($S$) of a transition-metal ion in a
crystalline solid can change with many factors, such as pressure,
strain, or temperature, to name a few. This phenomenon, known as
spin-state crossover, is of great importance in spintronics, as it
allows artificial control of magnetic properties of materials,
including coordination complexes with potential for molecular
switches \cite{TopCurrChem}. Not as widely known, spin-state
crossover also plays a crucial role in geophysics. A well studied
example is ferropericlase, (Mg,Fe)O, the second most abundant
mineral ($\sim20$ vol\%) in the the largest single region ($\sim55$
vol\%) of the Earth's interior -  the lower mantle. With increasing
pressure, ferrous iron (Fe$^{2+}$) in this mineral undergoes a
crossover from high-spin (HS) state, $S=2$, to low-spin (LS) state,
$S=0$, in the pressure range of 40-55 GPa \cite{MgFeO by Badro,
MgFeO by Lin 2005, MgFeO by Kantor 2006, MgFeO by Lin 2007, MgFeO by
Tsuchiya}. The intermediate-spin (IS) state, $S=1$, is not observed
in this mineral. The HS-LS crossover in ferropericlase directly
affects the structural, elastic, optical, and conducting properties
of this mineral \cite{MgFeO by Tsuchiya,MgFeO by Crowhurst,MgFeO by
Goncharov,MgFeO by Lin 2007b,Renata PNAS,MgFeO_by_Wu} and thus
affects mantle properties. \cite{Renata PNAS,Fp Pv by Lin,Fp Pv by
Hsu}.

\bigskip

In contrast, the spin-state crossover in iron-bearing magnesium
silicate (MgSiO$_3$) perovskite (Pv), the most abundant mineral
($\sim75$ vol\%) in the lower mantle, has been a source of
controversy for two main reasons. One is the coexisting ferrous and
ferric iron (Fe$^{3+}$) in this mineral with an imprecisely
estimated population ratio; the other is the lack of definitive
tools to directly probe iron spin state at high pressures. Two
techniques, X-ray emission spectroscopy (XES) and M\"ossbauer
spectroscopy, have been widely used, but their interpretation can be
ambiguous. The very similar XES spectra
\cite{XES-by-Badro,XES-by-Li} and M\"ossbauer spectra
\cite{QS-by-Jackson,QS-by-Li,QS-by-McCammon,QS-by-Lin} have been
interpreted in terms of HS-IS and HS-LS crossover in (Mg,Fe)SiO$_3$
Pv. Plenty of calculations on (Mg,Fe)SiO$_3$ Pv have been conducted
\cite{Hofmeister 2006, Zhang 2006, Stackhouse 2007, Bengtson 2008,
Koichiro, Site degeneracy Umemoto}, but consistency with experiments
was not achieved until very recently \cite{QS-by-Bengtson,
QS_MgFeSiO3_Hsu}. Now the spin state in (Mg,Fe)SiO$_3$ Pv is better
understood: the observed increase of iron nuclear quadrupole
splitting (QS) in M\"ossbauer spectra results from neither HS-IS nor
HS-LS crossover, but from the change in the $3d$ orbital occupancy
of the HS iron \cite{QS_MgFeSiO3_Hsu}. As to ferric iron in Pv,
possibly more abundant than ferrous iron ( Fe$^{3+}/\sum$Fe might be
as high as 2/3) \cite{Ferric_pop_McCammon,Ferric_pop_Frost}, its
spin-state crossover has remained unclear, as described below.

\bigskip

Previous experiments investigating the iron spin state in
aluminum-free MgSiO$_3$ Pv were focused mostly on ferrous iron
\cite{QS-by-Jackson,QS-by-McCammon}. Nevertheless, it was still
observed that the low concentration of ferric iron in the sample
exhibited an increase in QS with pressure, which suggests a
crossover from HS ($S=5/2$) to LS ($S=1/2$) state in the pressure
range of 30-70 GPa. In contrast, in Al-bearing samples, where ferric
iron occupies the dodecahedral (A) site, the QS remains unchanged up
to 100 GPa, which suggests the A-site iron remains in the HS state
\cite{QS-by-Li}. These results indicate that the ferric iron at the
octahedral (B) site undergoes a spin-state crossover. Such a
mechanism was recently confirmed by experiments using
(Mg$_{1-x}$Fe$_{x}$)(Si$_{1-x}$Fe$_{x}$)O$_3$ Pv ($x=0.1$) samples:
about half of the HS iron changes to LS state in the 45-60 GPa range
while the other half remain in the HS state all the way to 150 GPa
\cite{Ferric_by_Catalli}. So far, the computational studies on
(Mg$_{1-x}$Fe$_{x}$)(Si$_{1-x}$Fe$_{x}$)O$_3$ Pv have found a ground
state with HS iron at the A-site and LS iron at the B-site (A-HS;
B-LS) and an A-site HS-LS crossover that leads both A- and B-site
iron to a final LS state (A:LS; B-LS) at high pressures \cite{Zhang
2006,Stackhouse 2007}. These predictions are inconsistent with
experiments in two ways: (1) the predicted transition pressure is
too high; (2) the predicted HS iron concentration is too low.

\bigskip

\begin{figure}[pt]
\begin{center}
\includegraphics[
]{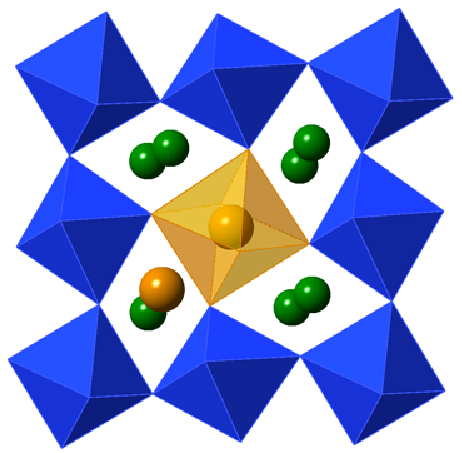}
\end{center}
\caption{} \label{atomic_structure}
\end{figure}

To compare with recent experiments \cite{Ferric_by_Catalli}, we
stabilize (Mg$_{1-x}$Fe$_{x}$)(Si$_{1-x}$Fe$_{x}$)O$_3$ Pv with
$x=0.125$ in all possible spin states using a 40-atom supercell
shown in Fig.~\ref{atomic_structure}. We also calculate the iron
nuclear electric field gradient (EFG) associated with each state, as
the nuclear hyperfine interaction has proven to be a unique
fingerprint to identify the spin states of transition-metal ions
\cite{QS_MgFeSiO3_Hsu,LaCoO3_EFG_Hsu}. The atomic structures were
fully optimized with damped variable cell shape molecular dynamics
\cite{VC-relax} implemented in the \textsc{quantum espresso} code
\cite{PWscf}, where the plane-wave pseudopotential method is adopted
\cite{PP}. These states were also independently confirmed via the
augmented plane-wave plus local orbitals (APW+lo) method
\cite{APW+lo} implemented in the WIEN2k code \cite{WIEN2k}, with
which the EFGs were calculated. The EFGs were converted to QSs with
$^{57}$Fe nuclear quadrupole moment $Q=0.16$ \cite{EFG using PAW}
and 0.18 barn for the possible uncertainty. To treat
(Mg$_{1-x}$Fe$_{x}$)(Si$_{1-x}$Fe$_{x}$)O$_3$ Pv, the density
functional theory plus Hubbard $U$ (DFT+$U$) method is necessary, as
standard DFT exchange-correlation functionals, the local density
approximation (LDA) and generalized gradient approximation (GGA),
sometimes lead to unwanted metallic states (especially at high
pressures), in which the iron spin states are not well defined.
Since the Hubbard $U$ of A- and B-site iron in each spin state is
unknown, we have to stabilize the desired spin state with a trial
$U$ and then extract the self-consistent $U$, referred to as
$U_{sc}$, using the linear response approach \cite{Cococcioni_LDA+U}
in a recently developed iterative procedure. This procedure is
equivalent to, but more efficient than the one published earlier
\cite{Usc_by_Kulik}, and has been successfully implemented
\cite{Campo10}. More details are described in the EPAPS
\cite{EPAPS}.

\bigskip

Within DFT+$U$, several combinations of iron spin states can be
stabilized. The A-site ferric iron can be stabilized in HS, IS, and
LS states. The B-site ferric iron can be stabilized not only in LS
state, but also in HS state that has not found in previous
calculations \cite{Zhang 2006,Stackhouse 2007}. The spin moments of
the A- and B-site iron can be either parallel or anti-parallel. The
$U_{sc}$ of ferric iron in Pv, listed in Table I, mainly depends on
the iron spin state, slightly depends on the occupied site, and
barely depends on pressure and alignment of spin moments.

\bigskip

\begin{figure}[pt]
\begin{center}
\includegraphics[
]{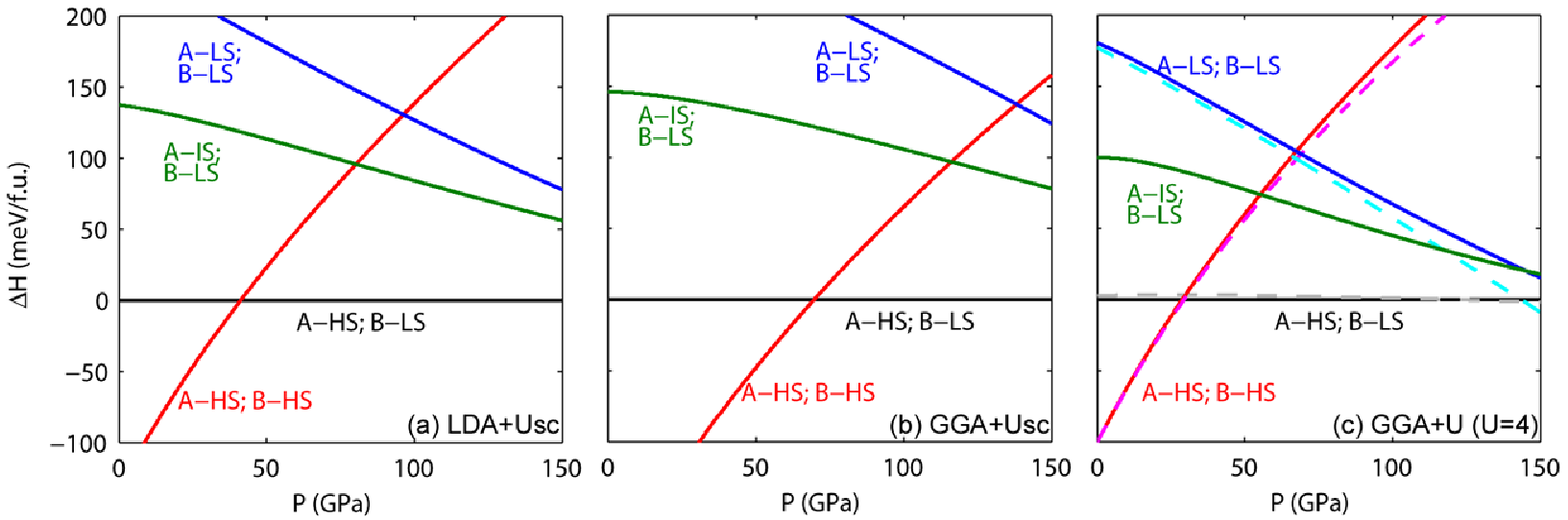}
\end{center}
\caption{} \label{delta_H}
\end{figure}

The relative enthalpy ($\Delta H$) of each stabilized state is shown
in Fig.~\ref{delta_H}, where the previously perceived ground state
(A-HS; B-LS) \cite{Zhang 2006,Stackhouse 2007} is used as a
reference. Remarkably, the actual ground state of
(Mg,Fe)(Si,Fe)O$_3$ Pv has HS iron on both sites (A-HS; B-HS),
regardless of the choice of exchange-correlation functional (LDA or
GGA) and Hubbard $U$ ($U_{sc}$ or 4 eV). These choices do not affect
the spin-state crossover either: an HS-LS crossover only occurs in
the B-site iron, while the A-site iron remains HS. As expected, the
predicted transition pressure ($P_{T}$) depends on the
exchange-correlation functional and Hubbard $U$: with LDA+$U_{sc}$,
$P_{T}=41$ GPa; with GGA+$U_{sc}$, $P_{T}=70$ GPa; with GGA+$U$
($U=4$ eV), $P_{T}=29$ GPa. (Coordination complexes also show
similar dependence \cite{Swart,Fouqueau}.) Notably, the alignment of
iron spins (parallel or anti-parallel), barely affects $P_{T}$, as
shown in Fig.~\ref{delta_H}(c). The $P_{T}$ predicted by
LDA+$U_{sc}$ and GGA+$U_{sc}$ best agree with the $P_{T}$ observed
in M\"ossbauer spectra, 50-60 GPa \cite{Ferric_by_Catalli}. The
LDA+$U_{sc}$ electronic density of states (DOS) of the two relevant
states (A-HS; B-HS and A-HS; B-LS) can be found in EPAPS
\cite{EPAPS}.

\bigskip

\begin{figure}[pt]
\begin{center}
\includegraphics[
]{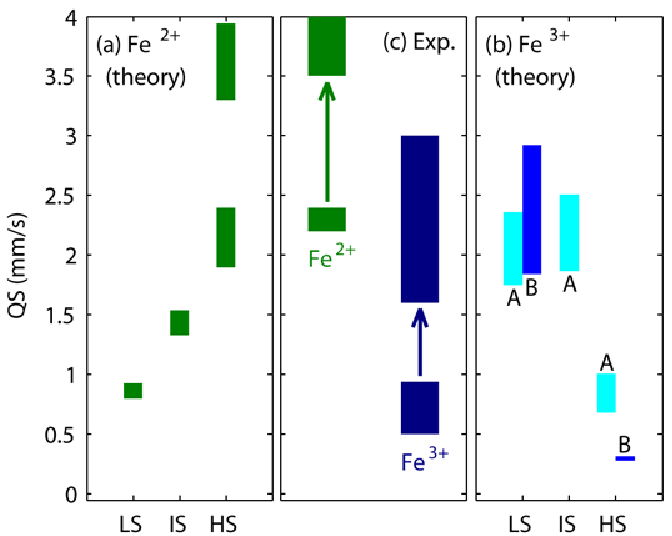}
\end{center}
\caption{} \label{QS}
\end{figure}

The calculated QSs of ferric iron (A- and B-site) and ferrous iron
(A-site) \cite{QS_MgFeSiO3_Hsu} in various spin states, along with
the measured QSs
\cite{QS-by-Jackson,QS-by-McCammon,Ferric_by_Catalli}, are shown in
Fig.~\ref{QS}. Clearly, our calculations on ferrous and ferric iron
in Pv are consistent with M\"ossbauer spectra. The HS-LS crossover
in the B-site ferric iron also helps to explain the decrease in the
XES satellite peak (K$\beta$') intensity
\cite{XES-by-Badro,XES-by-Li}. Interestingly, the QS of ferrous and
ferric iron exhibit exactly the opposite trends with respect to the
spin moment. This can be understood via their orbital occupancies.
The LS ferrous iron, although occupying the A site, is effectively
located near the center of a Fe-O octahedron, as it is vertically
displaced from the mirror plane \cite{Koichiro}. Its six $3d$
electrons doubly occupy the three orbitals with t$_{2g}$ character
and form a charge density with cubic-like shape \cite{Koichiro},
which barely contributes to the EFG and leads to a very small QS.
The HS ferric iron also has a small EFG (and thus QS), irrespective
of A or B site. This is because its five $3d$ electrons (all
spin-up) occupy all $3d$ orbitals, forming an almost spherically
shaped electron charge distribution that leads to a small EFG (and
thus QS). Similarly, the spin-up electrons in HS ferrous and LS
ferric iron barely contribute to EFG, as their charge distributions
are nearly spherical and cubic, respectively. It is their spin-down
electrons that contribute to the EFGs and lead to larger QSs. This
is why the spin moments of ferrous and ferric iron appear to affect
the QSs in an opposite manner.

\bigskip

\begin{figure}[pt]
\begin{center}
\includegraphics[
]{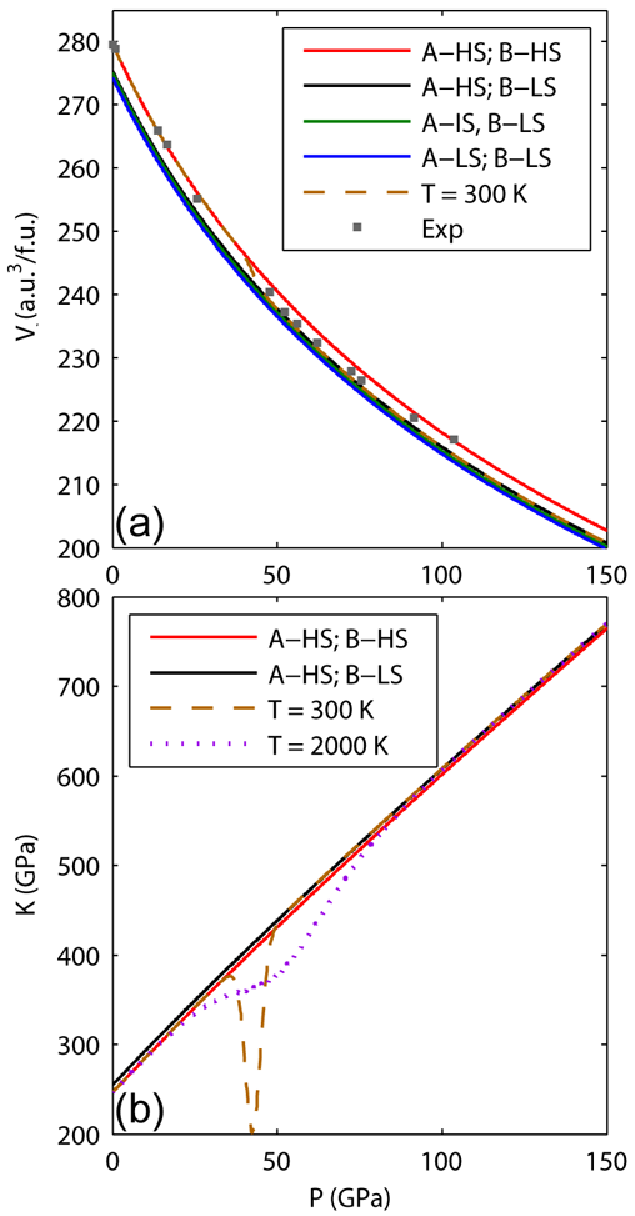}
\end{center}
\caption{} \label{VK_vs_PT}
\end{figure}

The LDA+$U_{sc}$ compression curves and bulk modulus
($K\equiv-VdP/dV$) of (Mg$_{1-x}$Fe$_{x}$)(Si$_{1-x}$Fe$_{x}$)O$_3$
Pv ($x=0.125$) along with the experimental data ($x=0.1$)
~\cite{Ferric_by_Catalli} are shown in Fig.~\ref{VK_vs_PT}. At low
pressures ($<45$ GPa), the experimental data falls on the calculated
compression curve corresponding to the (A-HS; B-HS) state. Starting
from $\sim$45 GPa, the data points deviate from the (A-HS; B-HS)
curve and then join the (A-HS; B-LS) curve at $\sim$60 GPa. Starting
from $\sim$100 GPa, the data deviates from the curve again. This,
however, is very likely to result from the questionable accuracy of
the Au pressure scale used in the experiment, as already discussed
in the case of (Mg,Fe)SiO$_3$ Pv \cite{U_in_Pv}. Notice that the
observed volume reduction further confirms the B-site HS-LS
crossover, as the previously perceived A-site HS-LS crossover barely
leads to a volume reduction, evident from the compression curves
(A-HS; B-LS and A-LS; B-LS) shown in Fig.~\ref{VK_vs_PT}(a). The
B-site spin-state crossover and the observed volume reduction in the
45-60 GPa range can be qualitatively understood via the Fe$^{3+}$
electronic configurations and Fe-O distances at A and B sites. With
all $3d$ orbitals occupied, HS iron has spherically-shaped electron
charge density and the largest radius compared with other spin
states, favoring longer Fe-O distances. Residing in the large
dodecahedral cage, the A-site iron can easily maintain longer Fe-O
distances and thus remain in HS state. In contrast, the Fe-O
octahedron has smaller size and shorter Fe-O distances. With
increasing pressure, the internal octahedron bond lengths can be
shortened enough to induce the HS-LS crossover. Since the $3d$
electrons of the B-site LS iron \textit{only} occupy the
$t_{2g}$-like orbitals pointing away from oxygen, the associated
Fe-O distances are significantly shorter than those of the HS iron
at the same pressure. Therefore, the spin change of the B-site iron
is accompanied by a noticeable octahedral (and thus unit-cell)
volume reduction. Such volume reduction leads to anomalous softening
in bulk modulus, as described below.

\bigskip

At finite temperatures, the spin-state crossover passes through a
mixed-spin (MS) state (namely, HS and LS coexist) within a finite
pressure range that increases with temperature. During the
crossover, the thermodynamic properties of the MS state exhibit
anomalous behavior that may affect mantle properties. One example is
the softening in bulk modules and its effect on the compressional
wave velocity, as already seen in ferropericlase \cite{MgFeO by
Crowhurst,Renata PNAS,MgFeO_by_Wu}. To estimate such anomaly in
(Mg,Fe)(Si,Fe)O$_3$ Pv, we employ a thermodynamic model similar to
that used in Ref.~\cite{Renata PNAS}. Here, we do not include
vibrational free energy, as it barely affects the magnitude of the
anomaly, slightly increases the transition pressure, and uniformly
decreases the bulk modulus, as shown in the case of ferropericlase
\cite{Renata PNAS,MgFeO_by_Wu}. Indeed, the calculated $V(P)$ curve
of (Mg,Fe)(Si,Fe)O$_3$ Pv in the MS state (using LDA+$U_{sc}$) at
room temperature (300 K), shown as the dashed line in
Fig.~\ref{VK_vs_PT}(a), exhibits a volume reduction ($\sim$1.2\%)
around the predicted $P_T$, 41 GPa. This reduction leads to a
significant softening in bulk modulus, as shown in
Fig.~\ref{VK_vs_PT}(b). The softening is still prominent at 2000 K,
the temperature near the top of the lower mantle ($\sim$660 km
deep). Given the abundance of iron-bearing Pv and the possibly high
population of ferric iron, this softening may have a noticeable
impact on the mantle properties, including possible anomalies in the
seismic wave velocities.

\bigskip

In summary, with a series of DFT+$U$ calculations, we have shown
that the actual ground state of (Mg,Fe)(Si,Fe)O$_3$ perovskite has
high-spin ferric iron on both A and B sites. It is the B-site ferric
iron that undergoes a crossover from high-spin to low-spin state
with increasing pressure, while the A-site iron remains in the
high-spin state. The calculated quadrupole splittings and the
compression curves are consistent with experiments. The volume
reduction accompanying the B-site HS-LS crossover leads to a
significant softening in bulk modulus, which suggests a possible
source of seismic-velocity anomalies in the lower mantle. This work,
one more time, demonstrates that the nuclear hyperfine interaction,
combined with first-principles calculations, can be a useful tool to
identify the spin states of transition-metal ions in solids under
high pressures.

\bigskip

This work was primarily supported by the MRSEC Program of NSF under
Award Number DMR-0212302 and DMR-0819885, and partially supported by
EAR-081272, EAR-1047629 and ATM-0426757 (VLab). P.B. was supported
by the Austrian Science Fund (SFB F41, "ViCoM"). Calculations were
performed at the Minnesota Supercomputing Institute (MSI).

\bigskip

Table I. $U_{sc}$, the self-consistent Hubbard $U$ (in eV), of
ferric iron on the A and B site in each spin state.

\begin{tabular}{ccc}
\hline\hline & \ \ \ A site \ \ \  & \ \ \ B site \ \ \  \\ \hline \
HS $\left( S=5/2\right) $ \  & 3.7 & 3.3 \\ \hline IS $\left(
S=3/2\right) $ & 4.6 & $-$ \\ \hline LS $\left( S=1/2\right) $ & 5.2
& 4.9 \\ \hline\hline
\end{tabular}

\newpage

{\Large Figure Captions}

\bigskip

Fig. 1. (Color online) Atomic structure of
(Mg$_{0.875}$Fe$_{0.125}$)(Si$_{0.875}$Fe$_{0.125}$)O$_3$ Pv,
configured with the shortest iron-iron distance, viewing along the
[001] direction. Large (orange) and small (green) spheres represent
Fe and Mg sites, respectively. Si-O and Fe-O octahedra are
shown in opaque (blue) and translucent (orange) colors.

\bigskip

Fig. 2. (Color online) Relative enthalpies of
(Mg$_{0.875}$Fe$_{0.125}$)(Si$_{0.875}$Fe$_{0.125}$)O$_3$ Pv in
different spin states obtained using different functionals and
Hubbard $U$. The reference state has HS iron in the A site and LS
iron in the B site (A-HS; B-LS). Predicted transition pressures by
LDA+$U_{sc}$ (a), GGA+$U_{sc}$ (b), and GGA+$U$ with $U=4$ eV (c)
are 41 and 70, and 29 GPa, respectively. Dashed lines in (c)
correspond to anti-parallel spins at A- and B-sites.

\bigskip

Fig. 3. (Color online) Calculated QSs of (a) ferrous iron
\cite{QS_MgFeSiO3_Hsu} and (b) ferric iron in MgSiO$_3$ Pv. Letter A
and B in (b) refer to iron-occupying site. Arrows in (c) indicate
the measured effect of pressure on QSs
\cite{QS-by-Jackson,QS-by-McCammon,Ferric_by_Catalli}.

\bigskip

Fig. 4. (Color online) Compression curves (a) and bulk modulus (b)
of (Mg$_{1-x}$Fe$_{x}$)(Si$_{1-x}$Fe$_{x}$)O$_3$ Pv computed with
LDA+$U_{sc}$ ($x=0.125$) and room-temperature measurements ($x=0.1$)
\cite{Ferric_by_Catalli}. Both the measured and calculated
compression curves exhibit a clear reduction accompanying with the
B-site HS-LS crossover, which leads to a softening in bulk modulus
shown in (b).

\newpage

\newpage

\newpage

\newpage


\begin{thebibliography}{9}

\bibitem{TopCurrChem} \textit{Spin Crossover in Transitoin Metal Complexes
I-III}, Top. Curr. Chem. \textbf{233-235}, edited by P. G\"{u}tlich
and H. A. Goodwin (Springer, 2004).

\bibitem{MgFeO by Badro} J. Badro \textit{et al}., Science \textbf{300},
789 (2003).

\bibitem{MgFeO by Lin 2005} J.-F. Lin \textit{et al}., Nature \textbf{436},
377 (2005).

\bibitem{MgFeO by Kantor 2006} I. Kantor, L. S. Dubrovinsky, and C. A.
McCammon, Phys. Rev. B \textbf{73}, 100101(R) (2006).

\bibitem{MgFeO by Lin 2007} J.-F. Lin \textit{et al}., Science \textbf{317},
1740 (2007).

\bibitem{MgFeO by Tsuchiya} T. Tsuchiya \textit{et al}, Phys. Rev. Lett.
\textbf{96}, 198501 (2006).

\bibitem{MgFeO by Crowhurst} J. Crowhurst \textit{et al}., Science
\textbf{319}, 451 (2008).

\bibitem{MgFeO by Goncharov} A. F. Goncharov \textit{et al}., Science
\textbf{312}, 1205 (2006).

\bibitem{MgFeO by Lin 2007b} J.-F. Lin \textit{et al}., Geophys. Res. Lett.
\textbf{34}, L16305 (2007).

\bibitem{Renata PNAS} R. M. Wentzcovitch \textit{et al}., Proc. Natl. Acad.
Sci. \textbf{106}, 847 (2009).

\bibitem{MgFeO_by_Wu} Z. Wu \textit{et al}., Phys. Rev. B
\textbf{80}, 014409 (2009).

\bibitem{Fp Pv by Lin} J.-F. Lin and T. Tsuchiya, Phys. Earth Planet.
In. \textbf{170}, 248 (2008), and references therein.

\bibitem{Fp Pv by Hsu} Han Hsu \textit{et al}., Rev. Mineral Geochem.
\textbf{71}, 169 (2010), and references therein.

\bibitem{XES-by-Badro} J. Badro \textit{et al}., Science
\textbf{305}, 383 (2004).

\bibitem{XES-by-Li} J. Li \textit{et al}., Proc. Natl. Acad.
Sci. \textbf{101}, 14027 (2004).

\bibitem{QS-by-Jackson} J. M. Jackson \textit{et al}., Am. Mineral.
\textbf{90}, 199 (2005).

\bibitem{QS-by-Li} J. Li \textit{et al}., Phys. Chem. Minerals.
\textbf{33}, 575 (2006).

\bibitem{QS-by-McCammon} C. McCammon \textit{et al}., Nature Geosci.
\textbf{1}, 684 (2008).

\bibitem{QS-by-Lin} J.-F. Lin \textit{et al}., Nature Geosci.
\textbf{1}, 688 (2008).

\bibitem{Hofmeister 2006} A. M. Hofmeister, Earth Planet. Sci. Lett.
\textbf{243}, 44 (2006).

\bibitem{Zhang 2006} F. Zhang and A. R. Oganov, Earth Planet. Sci. Lett.
\textbf{249}, 436 (2006).

\bibitem{Stackhouse 2007} S. Stackhouse \textit{et al}., Earth Planet.
Sci. Lett. \textbf{253}, 282 (2007).

\bibitem{Bengtson 2008} A. Bengtson, K. Persson, and D. Morgan, Earth
Planet. Sci. Lett. \textbf{265}, 535 (2008).

\bibitem{Koichiro} K. Umemoto \textit{et al}., Earth Planet. Sci. Lett.
\textbf{276}, 198 (2008).

\bibitem{Site degeneracy Umemoto} K. Umemoto, H. Hsu, and R. M.
Wentzcovitch, Phys. Earth Planet. In. \textbf{180}, 209 (2010).

\bibitem{QS-by-Bengtson} A. Bengtson \textit{et al}., Geophys. Res. Lett.
\textbf{36}, L15301 (2009).

\bibitem{QS_MgFeSiO3_Hsu} Han Hsu \textit{et al}., Earth Planet. Sci. Lett.
\textbf{294}, 19 (2010).

\bibitem{Ferric_pop_McCammon} C. McCammon, Nature \textbf{387}, 694
(1997).

\bibitem{Ferric_pop_Frost} D. Frost \textit{et al}., Nature
\textbf{428}, 409 (2004).

\bibitem{Ferric_by_Catalli} K. Catalli \textit{et al}., Earth Planet.
Sci. Lett. \textbf{289}, 68 (2010).

\bibitem{LaCoO3_EFG_Hsu} Han Hsu \textit{et al}., Phys. Rev. B \textbf{82},
100406(R) (2010).

\bibitem{VC-relax} R. M. Wentzcovitch, J. L. Martins, and G. D. Price,
Phys. Rev. Lett. \textbf{70}, 3947 (1993).

\bibitem{PWscf} P. Giannozzi \textit{et al}., J. Phys.: Condens. Matter
\textbf{21}, 395502 (2009).

\bibitem{PP} The pseudopotentials used in this work are the same as
those in Ref.~\cite{Koichiro}.

\bibitem{APW+lo} G. Madsen \textit{et al}., Phys. Rev. B \textbf{64},
195134 (2001).

\bibitem{WIEN2k} P. Blaha \textit{et al}., \textit{WIEN2k, An Augmented
Plane Wave Plus Local Orbitals Program for Calculating Crystal
Properties}, edited by K. Schwarz, Techn. Universit\"{a}t Wien,
Vienna (2001).

\bibitem{EFG using PAW} H. M. Petrilli \textit{et al}., Phys. Rev. B
\textbf{57}, 14690 (1998).

\bibitem{Cococcioni_LDA+U} M. Cococcioni and S. de Gironcoli, Phys. Rev.
B \textbf{71}, 035105 (2005).

\bibitem{Usc_by_Kulik} H. Kulik \textit{et al}., Phys. Rev. Lett.
\textbf{97}, 103001 (2006).

\bibitem{Campo10} V. L. Campo Jr and M. Cococcioni, J. Phys.: Condens. Matter
\textbf{22}, 055602 (2010).

\bibitem{EPAPS} See EPAPS.

\bibitem{Swart} M. Swart \textit{et al}, J. Phys. Chem. A
\textbf{108}, 5479 (2004).

\bibitem{Fouqueau} A. Fouqueau \textit{et al}, J. Chem. Phys.
\textbf{122}, 044110 (2005).

\bibitem{U_in_Pv} Han Hsu \textit{et al}., Phys. Earth Planet. In.
(2011), doi:10.1016/j.pepi.2010.12.001

\end{thebibliography}
\end{document}